\documentclass[11pt,a4paper,useAMS,usenatbib]{article}
\usepackage{jcappub}
\usepackage[english]{babel}

\usepackage{fancyhdr}
\usepackage{amsfonts}
\usepackage{amsmath}
\usepackage{amssymb}
\usepackage{multicol}
\usepackage{layout}
\usepackage{graphicx}
\usepackage{epstopdf}

\usepackage{times}
\usepackage{natbib}

\newif\ifAMStwofonts
\AMStwofontstrue

\title{Constraining Primordial Magnetic Fields with Future Cosmic Shear Surveys}

\author[a]{C. Fedeli}
\author[b,c,d]{and L. Moscardini}

\affiliation[a]{Department of Astronomy, University of Florida, 211 Bryant Space Science Center, \\Gainesville, FL 32611, USA}
\affiliation[b]{Dipartimento di Astronomia, Universit\`a di Bologna, Via Ranzani 1, 40127 Bologna, Italy}
\affiliation[c]{INAF - Osservatorio Astronomico di Bologna, Via Ranzani 1, 40127 Bologna, Italy}
\affiliation[d]{INFN, Sezione di Bologna, Viale Berti-Pichat 6/2, 40127 Bologna, Italy}

\emailAdd{cosimo.fedeli@astro.ufl.edu}

\abstract{The origin of astrophysical magnetic fields observed in galaxies and clusters of galaxies is still unclear. One possibility is that primordial magnetic fields generated in the early Universe provide seeds that grow through compression and turbulence during structure formation. A cosmological magnetic field present prior to recombination would produce substantial matter clustering at intermediate/small scales, on top of the standard inflationary power spectrum. In this work we study the effect of this alteration on one particular cosmological observable, cosmic shear. We adopt the semi-analytic halo model in order to describe the non-linear clustering of matter, and feed it with the altered mass variance induced by primordial magnetic fields. We find that the convergence power spectrum is, as expected, substantially enhanced at intermediate/small angular scales, with the exact amplitude of the enhancement depending on the magnitude and power-law index of the magnetic field power spectrum. Specifically, for a fixed amplitude, the effect of magnetic fields is larger for larger spectral indices. We use the predicted statistical errors for a future wide-field cosmic shear survey, on the model of the ESA Cosmic Vision mission \emph{Euclid}, in order to forecast constraints on the amplitude of primordial magnetic fields as a function of the spectral index. We find that the amplitude will be constrained at the level of $\sim 0.1$ nG for $n_B\sim -3$, and at the level of $\sim 10^{-7}$ nG for $n_B\sim 3$. The latter is at the same level of lower bounds coming from the secondary emission of gamma-ray sources, implying that for high spectral indices \emph{Euclid} will certainly be able to detect primordial magnetic fields, if they exist. The present study shows how large-scale structure surveys can be used for both understanding the origins of astrophysical magnetic fields and shedding new light on the physics of the pre-recombination Universe.}
\keywords{weak gravitational lensing; primordial magnetic fields.}

\begin{document}
\maketitle

\section{Introduction}\label{sct:introduction}

Magnetic fields are ubiquitous in the Universe. They are found, with varying intensities, in all kinds of astrophysical objects, ranging from planets ($B\sim 1$~G) to massive clusters of galaxies ($B \sim 1~\mu$G, see \cite{WI02.1} for a review). The origin of astrophysical magnetic fields is still debated, and two main scenarios have been devised. According to the first, magnetic fields form inside stars through mechanism such as the Biermann battery \cite{BI50.1}, they are then amplified due to dynamo mechanisms \cite{VA72.1,VA72.2,BE96.1}, and are finally diffused at larger scales thanks to supernova winds and Active Galactic Nuclei (AGN) feedback. The second scenario postulates instead that magnetic fields of very low intensity were generated in the early Universe, during inflation \cite{TU88.1,RA92.1} or phase transitions \cite{QU89.1} (for recent reviews see \cite{KA11.1,YA12.1}), and were subsequently amplified during gas infall within dark matter halos due to the conservation of the magnetic flux and the onset of turbulence \cite{SC12.1}. One interesting aspect of this latter scenario is that magnetic fields present in the early Universe are capable of generating and amplifying density perturbations in the baryonic fluid through the Lorentz force. Thanks to the gravitational coupling, these density perturbations would propagate to the dark matter fluid as well. In fact, for some time it was believed that such magnetic fields could be responsible for the generation of the seed density fluctuations for structure formation \cite{WA78.1}, until it was realized that the resulting power spectrum would be totally different from the observed one \cite{KI96.1,SU98.2}. Still, if Primordial Magnetic Fields (PMFs henceforth) existed, they would have altered the distribution of density fluctuations laid down during inflation.

The presence of magnetic fields in the early Universe has been considered in the literature as early as in \cite{ZE65.1,TH67.1,RE72.1}. Subsequently, the impacts of such fields on the primordial nucleosynthesis and the formation of the Cosmic Microwave Background (CMB henceforth) have been thoroughly studied \cite{KO00.1}, and constraints on their strength given by the observed elemental abundance \cite{CH94.1,CH96.1,GR96.1,KE96.1,KA12.1} and CMB temperature power spectrum \cite{BA97.1,SU98.1,JE00.1,SU02.1,YA10.1,YA10.2,YA10.3} have been put (see also \cite{YA12.2} for forecasted PMF bounds from \emph{Planck}). These constraints are in general compatible with a magnetic field strength, extrapolated at present time, that is $\lesssim 1$nG. More recent works studied the impact of PMFs of this level on several aspects of structure formation, such as the formation of voids \cite{DE97.1}, the redshift-space correlation function of galaxies \cite{GO03.1,SE03.3}, cosmic reionization \cite{SE05.2,TA06.2}, and $21$-cm fluctuations \cite{TA06.3}.

The only direct way of measuring the strength of magnetic fields on cosmological scales is the Faraday rotation that these fields induce on the polarization direction of high-redshift sources, such as galaxies and quasars. An early example of such a measurement can be found in Vall\'ee \cite{VA90.1}, who reported an upper limit on the strength of cosmological magnetic fields of $\sim 0.3$~nG $\left(0.023/\Omega_{\mathrm{b},0}h^2\right)$, assuming that all baryons are completely ionized in the low-redshift Universe (note that the author actually quotes a lower value, which is however based on an electron number density substantially higher than today's commonly accepted value). Earlier works (e.g. \cite{BR70.1,VA83.1}) as well as the subsequent review by Kronberg \cite{KR94.1} seemed however to point toward values somewhat higher than this. In fact, \cite{KO98.1} found an upper limit on the magnetic mean-field amplitude of $\sim 0.5~$nG $\left( 0.023/\Omega_{\mathrm{b},0}h^2\right)$ on scales $\sim 10-50~h^{-1}$ Mpc, while \cite{BL99.1} reported findings varying from $\sim1-10~$nG, depending on the coherence length they considered. More recently \cite{XU06.1} found an upper limit of $\sim 0.8~\mu$G for the intergalactic magnetic field strength on a scale of $\sim 1~h^{-1}$ Mpc.

Several more exotic possibilities for probing cosmological magnetic fields exist. For instance, in \citep{KO96.1} the authors have proposed to use the CMB itself as a source of polarized radiation in order to measure the Faraday rotation due to PMFs. At around the same time, Plaga \cite{PL95.1} put forward the possibility to use the secondary emission of gamma-ray sources such as gamma-ray bursts (see also \cite{IC08.1}) or blazars in order to put constraints on the strength of intergalactic magnetic fields. As it turns out, the PMF amplitude should be at least of $\sim 10^{-8}-10^{-6}$ nG in order to explain this kind of observations \cite{NE10.1,TA11.1,TA11.2}.

In the present paper we consider a novel way to investigate PMFs. As mentioned above, cosmological magnetic fields present before recombination are expected to alter the large-scale matter distribution. Specifically, PMFs introduce substantial power on top of the inflationary one at intermediate/small scales. Cosmic shear, that is the weak gravitational deflection of light by the Large-Scale Structure (LSS) should be ideally suited to probe those scales. Here we compute the cosmic shear power spectrum produced in cosmological models with PMFs having a variety of spectral slopes. Then, using the specifications of future wide-field weak lensing surveys on the model of \emph{Euclid}\footnote{http://www.euclid-ec.org/} \cite{LA11.1}, a recently accepted ESA Cosmic Vision program mission, we estimate the statistical error on the power spectrum and hence derive forecasted constraints on the amplitude of PMFs. A similar study has been recently performed in \cite{PA12.1}, which we discuss in Section \ref{sct:conclusions}.

The rest of the paper is organized as follows. In Section \ref{sct:linear} we review the impact of PMFs on the linear stage of structure formation, with particular care to the growth factor of magnetically-induced matter density fluctuations and their linear power spectrum. In Section \ref{sct:nonlinear} we show how the modifications at the linear level can be mapped to the non-linear level, describing the computation of the mass function, halo bias, and eventually the non-linear matter power spectrum in presence of PMFs. In Section \ref{sct:shear} we turn attention to the convergence power spectrum and how it gets modified by PMFs, thus giving forecasted constraints on the amplitude of PMFs as a function of their spectral index. In Section \ref{sct:conclusions} we summarize our conclusions.

In this work we adopted the cosmological parameters suggested by the latest WMAP data release in conjunction with Baryon Acoustic Oscillation (BAO) and Hubble constant measurements \cite{KO11.1}. This implies a matter density parameter of $\Omega_{\mathrm{m},0} = 0.272$, a dark energy density parameter of $\Omega_{\Lambda,0} = 1-\Omega_{\mathrm{m},0}$, a baryon density parameter of $\Omega_{\mathrm{b},0} = 0.046$, and a Hubble constant of $H_0 = h~100$ km s$^{-1}$ Mpc$^{-1}$, with $h = 0.704$. The amplitude of primordial curvature perturbations is set to $\Delta_\mathcal{R}^2 = 2.44\times 10^{-9}$, corresponding to $\sigma_8 = 0.809$ in absence of PMFs, while the spectral index is given by $n = 0.963$. Note that PMFs have impacts on the CMB power spectrum as well. These display themselves both on scales smaller than those probed by WMAP, due to vortical perturbations \cite{SE01.1,SU02.1}, and on very large angular scales, due to anisotropic magnetic stresses \cite{DU00.1,MA02.1}. While the former do not affect the WMAP bounds on cosmological parameters \cite{YA12.1}, the latter, being at the percent level, might alter the inferred amplitude and slope of the inflationary curvature power spectrum. Awaiting for more detailed studies on this topic, in this work we chose to stick with the WMAP cosmological parameters. If not specified otherwise, we shall assume natural units such that the speed of light is $c = 1$ throughout.

\section{Linear structure growth}\label{sct:linear}

\subsection{Magnetic fields}\label{sct:pmfs}

PMFs affect the growth of density perturbations in the baryon fluid. Since density fluctuations in the dark matter component are sourced also by those in the baryonic component, magnetic fields indirectly affect the growth of dark matter perturbations as well. Let us indicate with $\boldsymbol B(\boldsymbol x, t)$ the magnetic field that is immersed in the baryon fluid. For the sake of completeness, it is instructive to review the equations describing the evolution of small comoving perturbations for a magnetized fluid in an expanding cosmological background \citep{WA78.1}. While the continuity and Poisson equations are identical to the standard ones, the Euler equation displays a significant difference,

\begin{equation}
\frac{\partial \boldsymbol v(\boldsymbol x, t)}{\partial t} + H(t) \boldsymbol v(\boldsymbol x, t) = -\frac{\nabla\psi(\boldsymbol x, t)}{a(t)} + \boldsymbol S(\boldsymbol x, t)~,
\end{equation}
where

\begin{equation}
\boldsymbol S(\boldsymbol x, t) = \frac{\left[\nabla \times \boldsymbol B (\boldsymbol x, t)\right] \times \boldsymbol B (\boldsymbol x, t)}{4\pi a(t)\rho_\mathrm{b}(t)}~.
\end{equation}
In the previous equations we have neglected the role of pressure gradients, since these are irrelevant on scales much larger than the thermal Jeans length. It appears that the peculiar velocity of baryon density perturbations is not only sourced by potential perturbations, but also by magnetic fields through the Lorentz force.

Magnetic fields in turn obey the standard solenoidal condition $\nabla \cdot \boldsymbol B(\boldsymbol x, t) = 0$, and also they are fueled by the velocity fields via magnetic induction, through a backreaction mechanism that for a fluid with infinite conductivity reads

\begin{equation}
\frac{\partial}{\partial t} \left[ a^2(t) \boldsymbol B(\boldsymbol x, t) \right] = \frac{\nabla\times \left[ \boldsymbol v(\boldsymbol x, t) \times a^2(t)\boldsymbol B(\boldsymbol x, t)\right]}{a(t)}~.
\end{equation}
On scales much larger than the magnetic Jeans length (see below) the backreaction given by the right-hand-side of the previous equation can be neglected at first order, thus leading to the constancy of $a^2(t) \boldsymbol B(\boldsymbol x, t)$, that is $a^2(t) \boldsymbol B(\boldsymbol x, t) = \boldsymbol B (\boldsymbol x, t_0) \equiv \boldsymbol B_0(\boldsymbol x)$. From this, we note the interesting fact that the magnetic energy density,

\begin{equation}
\rho_B(t) \equiv \frac{\langle B^2(\boldsymbol x, t) \rangle}{8\pi} = \frac{\langle B_0^2(\boldsymbol x) \rangle}{8\pi a^4(t)} = \frac{\rho_{B,0}}{a^4(t)}
\end{equation}
(where angular brackets represent ensemble averages), evolves in the same way as the radiation energy density in an expanding Universe \cite{KI96.1}.

One important ingredient in order to estimate the impact of PMFs on the growth of cosmic structures is their correlation function. For a statistically homogeneous and isotropic magnetic field the correlation function in Fourier space can be written as \citep*{KR67.1,KI96.1}

\begin{equation}\label{eqn:correlation}
\left\langle \hat B_i(\boldsymbol k,t) \hat B_j(\boldsymbol k',t) \right\rangle = \frac{(2\pi)^3}{2}\delta_\mathrm{D}(\boldsymbol k - \boldsymbol k')\left(\delta_{ij} -\frac{k_ik_j}{k^2} \right) P_B(k,t)~.
\end{equation}
It is often assumed that the power spectrum $P_B(k,t)$ of PMFs is a power law,

\begin{equation}
P_B(k,t) = A_B(t) k^{n_B}~,
\end{equation}
in a given range of scales, $k\in [k_\mathrm{min}, k_\mathrm{max}]$, and vanishes otherwise. For the sake of completeness, in this work we considered a wide range of spectral indices. However we bear in mind that constraints from the gravitational wave background have been claimed to prefer $n_B \lesssim 0$ \citep{CA02.1}, although these constraints are conditional on independent determinations of the effective number of neutrino species. Also, joint constraints coming from the CMB and the linear matter distribution favor $n_B\lesssim 0$ \cite{YA10.2,PA11.1}, however such constraints strongly depend on the priors assumed in order to break the degeneracy between the magnetic amplitude and spectral index.

The ultraviolet cutoff scale $k_\mathrm{max}$ is often interpreted as the scale at which magnetic fields are dissipated due to radiative viscosity around recombination. Also, the primordial magnetic fields are usually assumed to be correlated up to very large scales, so that, effectively, $k_\mathrm{min} = 0$. Before going with more detail in the specific value of $k_\mathrm{max}$ adopted in this work, we note that, analogously to what happens for density fluctuations, the normalization of the magnetic field power spectrum can be written in terms of the variance of the magnetic fields smoothed on a certain comoving scale $\lambda$. Assuming a sharp Fourier-space smoothing function of width $k = 2\pi/\lambda < k_\mathrm{max}$ we obtain

\begin{equation}
\sigma_B^2(\lambda,t) = \frac{1}{2\pi^2}\int_0^{2\pi/\lambda} k^2 \mathrm{d}k~P_B(k,t) = \frac{2A_B(t)}{3+n_B} \frac{(2\pi)^{1+n_B}}{\lambda^{3+n_B}}~.
\end{equation}
Note that some authors (e.g., \cite{MA02.1}) do not divide by the factor $2$ in Eq. (\ref{eqn:correlation}), so that for a fixed spectral amplitude their magnetic field variance is twice as large. This is a point to keep in mind when comparing constraints on the amplitude of PMFs obtained in different works. The result in the previous equation is valid only if $n_B > -3$, which is a common assumption in PMF studies. It follows that $\sigma_B(\lambda,t)$ is the magnetic comoving mean-field amplitude across the scale $\lambda$ at cosmic time $t$.

Now we want to give some more details about the ultraviolet cutoff $k_\mathrm{max}$. According to \cite{JE98.1,SU98.2}, the scale at which magnetic fields suddenly dissipate can be estimated as (see also \cite{MA02.1})

\begin{equation}
\frac{1}{k_\mathrm{max}^2} = \frac{\lambda_\mathrm{max}^2}{(2\pi)^2} = \frac{\sigma_B^2(\lambda_\mathrm{max},t_\mathrm{r})}{4\pi\rho_{\gamma,\mathrm{r}}\sigma_\mathrm{T}}\int_0^{t_\mathrm{r}} \frac{\mathrm{d}t}{a^2(t)n_e(t)}~.
\end{equation}
In the previous equation $\rho_{\gamma,\mathrm{r}}$ is the radiation energy density at recombination, $\sigma_\mathrm{T}$ is the Thomson scattering cross section, and $n_e(t)$ is the number density of free electrons before the recombination time $t_\mathrm{r}$. We recall that for a fully ionized plasma with standard primordial abundances $n_e(t) = 0.52~\rho_\mathrm{b}(t)/\mu m_\mathrm{p}$, with $\mu = 0.59$. Considering a generic length scale $\lambda < \lambda_\mathrm{max}$ we get

\begin{equation}
\frac{\lambda_\mathrm{max}^2}{(2\pi)^2} = \frac{\sigma_B^2(\lambda,t_\mathrm{r})}{4\pi\rho_{\gamma,\mathrm{r}}\sigma_\mathrm{T}}
\left(\frac{\lambda}{\lambda_\mathrm{max}}\right)^{3+n_B} \frac{\mu m_\mathrm{p}}{0.52~\rho_{\mathrm{b},0}}\int_0^{t_\mathrm{r}}a(t) \mathrm{d}t~.
\end{equation}
It is safe to assume that before recombination the Universe behaves as an Einstein-de Sitter universe, so that the integral in the previous equation can be easily written as

\begin{equation}
\int_0^{t_\mathrm{r}} a(t) dt = \int_0^{a_\mathrm{r}} \frac{\mathrm{d}a}{H(a)} = \frac{2 a_\mathrm{r}^{5/2}}{5 H_0\sqrt{\Omega_{\mathrm{m},0}}}~.
\end{equation}
Also, the radiation energy density and the magnetic field variance have the same temporal evolution, so that the ultraviolet cutoff scale can be estimated as

\begin{equation}
\lambda_\mathrm{max}^{5+n_B} = \frac{2}{5}\pi\frac{\sigma_{B,0}^2(\lambda)}{\rho_{\gamma,0}\sigma_\mathrm{T}}\frac{\lambda^{3+n}}{H_0\sqrt{\Omega_{\mathrm{m},0}}} \frac{\mu m_\mathrm{p}}{0.52~\rho_{\mathrm{b},0}} a_\mathrm{r}^{5/2}~.
\end{equation}
By placing the relevant numerical values where necessary, in particular $a_\mathrm{r} = 1/(1+z_\mathrm{r}) \simeq 9.083 \times 10^{-4}$, we obtain

\begin{equation}\label{eqn:kmax}
\left(\frac{\lambda_\mathrm{max}}{\lambda}\right)^{5+n_B} \simeq 1.319\times 10^{-3}~h^2 \left[\frac{\sigma_{B,0}(\lambda)}{1 ~\mathrm{n}G}\right]^2 \left( \frac{\lambda}{1h^{-1}\mathrm{Mpc}} \right)^{-2} \left( \frac{\Omega_{\mathrm{b},0}h^2}{0.023} \right)^{-1} \left( \frac{\Omega_{\mathrm{m},0}h^2}{0.135} \right)^{-1/2}~.
\end{equation}
As an example, if we assume a typical value for the present-day magnetic field dispersion $\sigma_{B,0}(\lambda) = 0.1$ nG for $\lambda = 1~h^{-1}$Mpc and $h=0.704$ we obtain $\lambda_\mathrm{max} \simeq 18.7~h^{-1}$kpc if $n_B=-2$, $\lambda_\mathrm{max} \simeq 91.8~h^{-1}$kpc if $n_B=0$, and $\lambda_\mathrm{max} \simeq 182~h^{-1}$kpc if $n_B=2$. These are the values of the ultraviolet cutoff that shall be used in the remainder of this paper. We note that $\lambda_\mathrm{max}$ is a rather small scale, hence any cosmologically relevant scale over which we wish to smooth the magnetic field itself would be larger than this.

Another important length scale for magnetized fluids is the magnetic Jeans length, $\lambda_B$. This is the comoving transition scale below which perturbations in the magnetized fluid stop growing, due to the fact that magnetic pressure gradients counteract the gravitational pull \citep{SU98.2}. This length scale has been derived for the first time in \cite{KI96.1} for a fully baryonic Universe. Later, the generalization to cosmological models dominated by dark matter was provided by \cite{SE05.2}. The magnetic Jeans length reads implicitly as

\begin{equation}
a(t)\lambda_B(t) = \frac{2}{5} \frac{\sigma_B(\lambda_B(t), t)}{\sqrt{G \rho_\mathrm{m}(t)\rho_\mathrm{b}(t)}} = \frac{2}{5}\frac{\sigma_B(\lambda,t)}{\sqrt{G \rho_\mathrm{m}(t)\rho_\mathrm{b}(t)}} \left[\frac{\lambda}{\lambda_B(t)}\right]^{(3+n_B)/2}~.
\end{equation}
If the dispersion of the magnetic field evolves as $\sigma_B(\lambda,t) \propto a^{-2}(t)$, this implies that the magnetic Jeans length is independent of time, so that we can define $\lambda_B(t) = \lambda_{B,0} \equiv \lambda_B$. It follows that

\begin{equation}\label{eqn:kb}
\left(\frac{\lambda_B}{\lambda}\right)^{5+n_B} \simeq 0.2298~h^2 \left[\frac{\sigma_{B,0}(\lambda)}{1~\mathrm{nG}}\right]^2 \left( \frac{\lambda}{1h^{-1}\mathrm{Mpc}} \right)^{-2} \left( \frac{\Omega_{\mathrm{b},0}h^2}{0.023} \right)^{-1} \left( \frac{\Omega_{\mathrm{m},0}h^2}{0.135} \right)^{-1}~.
\end{equation}
It is easily seen that the magnetic Jeans scale is much larger than the ultraviolet cutoff scale.

There is one further potentially important length scale for the problem at hand, that is the standard thermal Jeans length for baryons. As mentioned above, the magnetically induced density fluctuations in dark matter are sourced by their baryonic counterparts. On scales smaller than the thermal Jeans length the latter are washed out by thermal pressure, hence the former cannot grow either. Normally the thermal Jeans scale is much smaller than the magnetic one, and in fact it has been ignored in the vast majority of works on PMFs. However, it has recently been argued (see \cite{SE05.2,SE08.2,KA10.1}) that the small-scale dissipation of PMFs due to ambipolar diffusion and decaying turbulence provides an extra source of energy that is injected in the baryonic fluid. This additional energy would increase the temperature of the inter-galactic medium by several orders of magnitude at low redshifts, and hence increase substantially the thermal Jeans scale. The computations of \cite{SE08.2} show that the thermal Jeans scale can indeed become larger than the magnetic one if one considers the strengthening of magnetic fields due to baryonic matter compression in collapsing structures. At the linear level however, the increase in temperature would be much lower than that, and hence it would be safe to assume the thermal Jeans scale to be always smaller than or comparable to the magnetic one. This is the assumption that we adopted in the rest of the paper. Plus, the extra heating due to matter (and hence magnetic field lines) compression occurring inside the first non-linear structures is arguably compensated by the cooling due to the much more abundant H$_2$ \cite{SE08.2}.

\subsection{Time evolution of density fluctuations}

Since dark matter is not directly coupled to magnetic fields, the evolution of linear dark matter density perturbations is given by the standard equation

\begin{equation}\label{eqn:dm}
\ddot\delta_\bullet(\boldsymbol x, t) + 2 H(t) \dot\delta_\bullet(\boldsymbol x, t) - 4\pi G \left[ \rho_\bullet(t)\delta_\bullet(\boldsymbol x, t) +  \rho_\mathrm{b}(t)\delta_\mathrm{b}(\boldsymbol x, t) \right] = 0~.
\end{equation}
In other words, density fluctuations in both baryons, $\delta_\mathrm{b}(\boldsymbol x,t)$, and dark matter, $\delta_\bullet(\boldsymbol x,t)$, source the growth of the latter due to gravitational instability. On the other hand, baryons are coupled with magnetic fields, hence the evolution of their density fluctuations follow a different equation \cite{KI96.1}, 

\begin{equation}\label{eqn:b}
\ddot\delta_\mathrm{b}(\boldsymbol x, t) + 2 H(t) \dot\delta_\mathrm{b}(\boldsymbol x, t) - 4\pi G \left[ \rho_\bullet(t)\delta_\bullet(\boldsymbol x, t) +  \rho_\mathrm{b}(t)\delta_\mathrm{b}(\boldsymbol x, t) \right] = \frac{\nabla\cdot\boldsymbol S(\boldsymbol x, t)}{a(t)}~.
\end{equation}
In this case, the magnetic fields also provide an additional force that contributes to the linear growth of density perturbations. By following, e.g., \cite{GO03.1} we define the density fluctuation of the total matter fluid as

\begin{equation}
\rho_\mathrm{m}(t)\delta_\mathrm{m}(\boldsymbol x, t) \equiv \rho_\bullet(t)\delta_\bullet(\boldsymbol x, t) + \rho_\mathrm{b}(t)\delta_\mathrm{b}(\boldsymbol x, t)~,
\end{equation}
where $\rho_\mathrm{m} = \rho_\bullet + \rho_\mathrm{b}$. It easily follows that $\delta_\mathrm{m}(\boldsymbol x, t) = f_\bullet \delta_\bullet(\boldsymbol x, t) + f_\mathrm{b}\delta_\mathrm{b}(\boldsymbol x, t)$, where $f_\bullet = \Omega_{\bullet,0}/\Omega_{\mathrm{m},0}$ and $f_\mathrm{b} = \Omega_{\mathrm{b},0}/\Omega_{\mathrm{m},0}$. By replacing $\delta_\bullet(\boldsymbol x, t) = \delta_\mathrm{m}(\boldsymbol x, t)/f_\bullet - \delta_\mathrm{b}(\boldsymbol x, t)f_\mathrm{b}/f_\bullet$ into Eq. (\ref{eqn:dm}) and making use of Eq. (\ref{eqn:b}) we obtain

\begin{equation}\label{eqn:m}
\ddot\delta_\mathrm{m}(\boldsymbol x, t) + 2 H(t) \dot\delta_\mathrm{m}(\boldsymbol x, t) - 4\pi G \rho_\mathrm{m}(t)\delta_\mathrm{m}(\boldsymbol x, t) = f_\mathrm{b} \frac{\nabla\cdot\boldsymbol S(\boldsymbol x, t)}{a(t)}~.
\end{equation}

Eq. (\ref{eqn:m}) is what one needs to solve in order to determine the time evolution of density fluctuations in the presence of magnetic fields, and hence in order to find out the growth factor. This equation can be further simplified by recalling that on the scales $\lambda \gg \lambda_B$ of our interest the magnetic field evolves simply as $\boldsymbol B(\boldsymbol x, t) = \boldsymbol B_0(\boldsymbol x)/a^2(t)$, so that the forcing term of Eq. (\ref{eqn:m}) can be rewritten as $f_\mathrm{b} \nabla\cdot \boldsymbol S_0(\boldsymbol x)/a^3(t)$. Now, the homogeneous part of the previous equation is the standard equation for the evolution of density fluctuations on large scales. Its solution can be written as a linear combination of the usual growing and decaying modes of structure growth, $D_+(t)$ and $D_-(t)$. In order to obtain the most general solution of Eq. (\ref{eqn:m}) we need to find a particular solution of the same equation and sum it to the general solution of the homogeneous equation. It follows that the solution is

\begin{equation}
\delta_\mathrm{m}(\boldsymbol x, t) = \Gamma_+(\boldsymbol x)D_+(t) + \Gamma_-(\boldsymbol x)D_-(t) + f_\mathrm{b} \nabla\cdot \boldsymbol S_0(\boldsymbol x) M(t)\simeq \Gamma_+(\boldsymbol x)D_+(t) + f_\mathrm{b} \nabla\cdot \boldsymbol S_0(\boldsymbol x) M(t)~.
\end{equation}
The function $M(t)$ represents the temporal evolution of the matter density fluctuations induced by PMFs, and it is a solution of the differential equation

\begin{equation}
\ddot M(t) + 2H(t)\dot M(t) - 4\pi G\rho_\mathrm{m}(t) M(t) = \frac{1}{a^3(t)}~.
\end{equation}
Such a solution has to be found through numerical integration for a $\Lambda$CDM Universe. At very early times however the Universe is well represented by an Einstein-de Sitter model, for which the previous equation admits an analytic solution for $t\ge t_\mathrm{r}$ (before recombination magnetically induced density perturbations cannot grow because of the tight coupling between baryons and photons), that is \cite{KI96.1}

\begin{equation}\label{eqn:eds}
M(t) = t_0^2 \left[ \frac{9}{10} \left( \frac{t}{t_\mathrm{r}} \right)^{2/3} + \frac{3}{5}\left(\frac{t}{t_\mathrm{r}}\right)^{-1} - \frac{3}{2}\right]\simeq  t_0^2 \frac{9}{10} \left( \frac{t}{t_\mathrm{r}} \right)^{2/3}~.
\end{equation}
We used this solution in order to set up the initial conditions for the numerical integration of the differential equation. It is interesting to note that the growing part of the analytic solution above is identical to the standard growth factor in an Einstein-de Sitter universe. By numerically integrating the differential equation we verified that this the case also for the more general $\Lambda$CDM case.

\begin{figure}
\centering
\includegraphics[width=0.65\hsize]{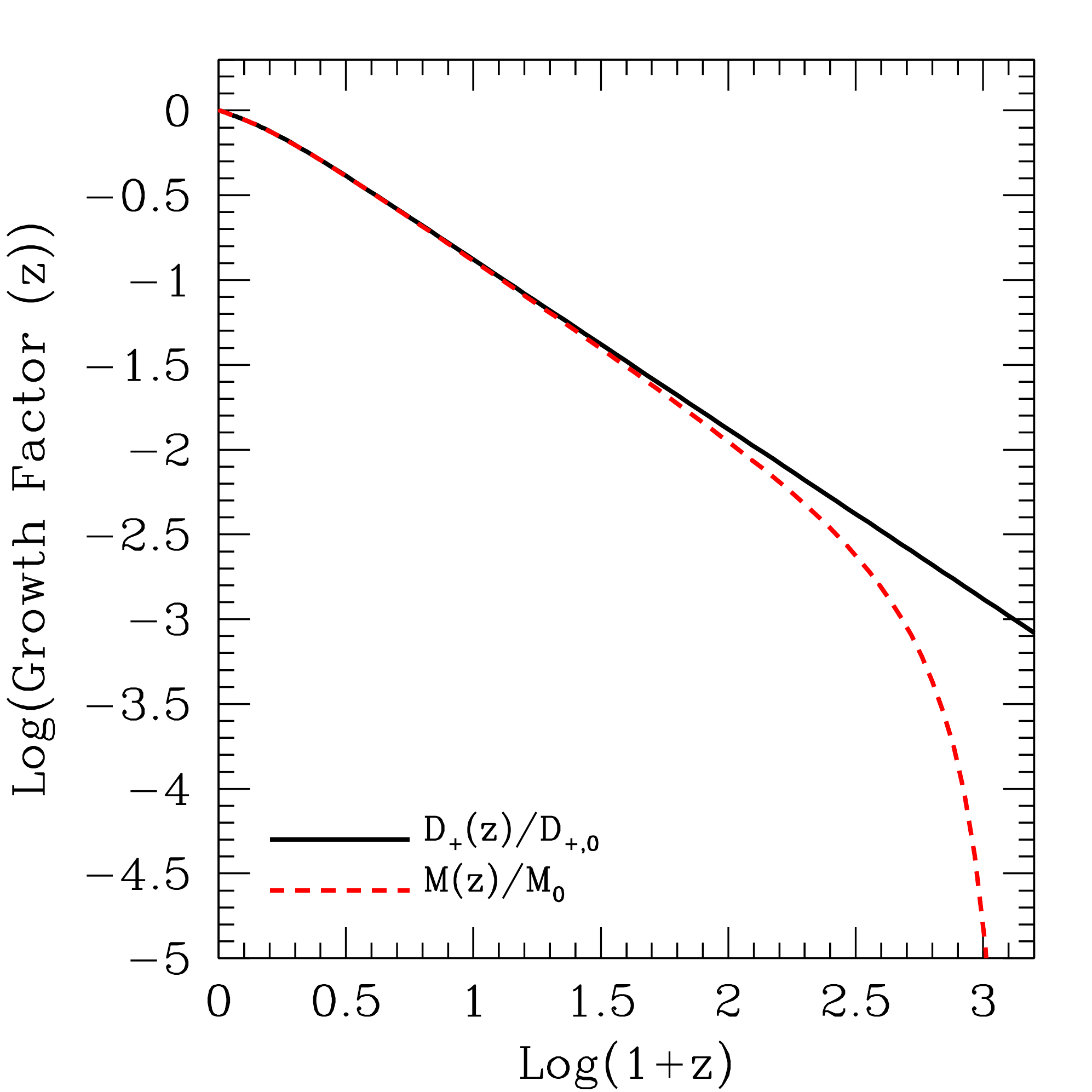}
\caption{The growth factors relevant for the present work, normalized to unity at $z=0$. The black solid line represents the growth factor of standard inflationary density perturbations, while the red dashed line shows the growth factor of density perturbations induced by PMFs.}
\label{fig:growth}
\end{figure}

\begin{figure}
\centering
\includegraphics[width=0.65\hsize]{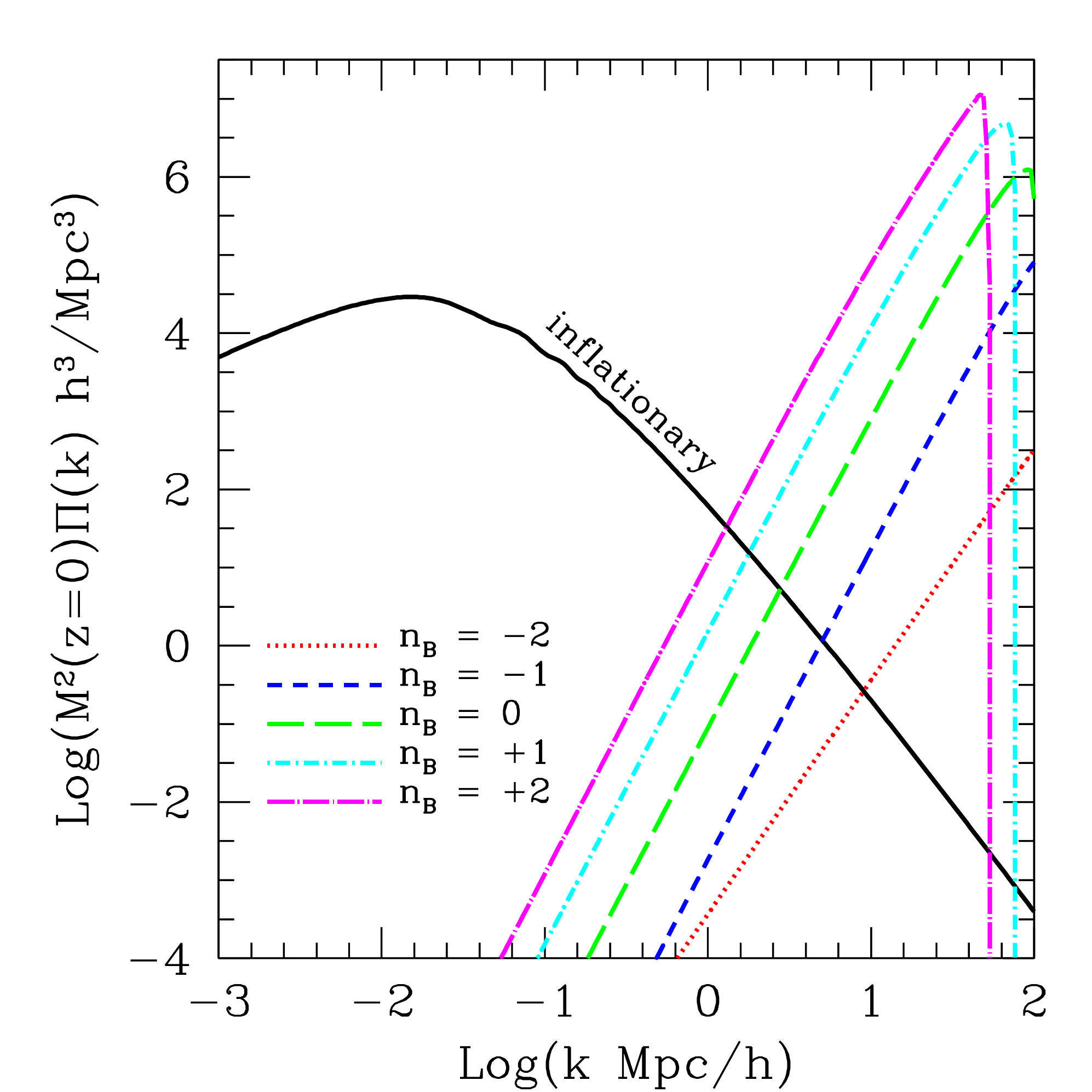}
\caption{The power spectrum of density fluctuations induced by PMFs. Each line style and color refer to a different magnetic spectral index, as labeled, while the spectral amplitude is set by $\sigma_{B,0}(1~h^{-1}\mathrm{Mpc}) = 0.1~\mathrm{nG}$. The black solid line shows the power spectrum of primordial density fluctuations for reference.}
\label{fig:spectrum}
\end{figure}

In Figure \ref{fig:growth} we compare the standard growth factor $D_+(z)$ for inflationary density perturbations with the growth factor of density perturbations induced by PMFs, $M(z)$. At low redshift the two growth factors coincide, as is the case for the Einstein-de Sitter universe. In fact, according to Eq. (\ref{eqn:eds}), $M(z) \propto (1+z)^{-1}$ once the non-growing part has been dissipated. This is due to the fact that, once the density fluctuations have been generated by magnetic fields, they grow due to gravitational instability as the primordial ones. The main difference between the two function stays at high redshift, and in particular in the fact that, while magnetically induced density perturbations cannot grow before recombination (Log$(1+z_\mathrm{r})\simeq 3$), the standard density perturbations have been growing for a while by $z_\mathrm{r}$, so that the decaying mode has already dissipated. As a consequence, we have that $M(z)\rightarrow 0$ as $z\rightarrow z_\mathrm{r}$. Note that in the redshift range we shall be interested in this work ($z\lesssim 5$), the two growth factors can be effectively considered as identical.

\subsection{Matter power spectrum}\label{sct:linearspectrum}

In order to evaluate the linear matter power spectrum in presence of PMFs we make the assumption that primordial density fluctuations are uncorrelated from the density fluctuations induced by magnetic fields. This is an assumption that is often employed in the literature, although the presence or absence of such correlations depends on the mechanism responsible for the generation of PMFs, over which a consensus has not been reached yet (see \cite{YA06.1} for a discussion). If however that is the case, then it is easily seen that the total matter power spectrum is

\begin{equation}\label{eqn:spectrum}
P_\mathrm{m}(k,t) = D^2_+(t) P(k) + M^2(t) \Pi(k)~,
\end{equation}
where $P(k)$ is the standard matter power spectrum, while $\Pi(k)$ is the power spectrum of matter density fluctuations induced by magnetic fields. In writing down the previous equation we assumed that the decaying mode has dissipated, so that the growth of standard density perturbations is effectively given only by the growing mode. It has been shown in the literature \cite{KI96.1} that the power spectrum of magnetically induced fluctuations can be written as

\begin{equation}
\Pi(k) = \frac{f_\mathrm{b}^2}{\left(4\pi\rho_{\mathrm{b},0}\right)^2}\int _0^{+\infty}\mathrm{d}q\int_{-1}^1\mathrm{d}\mu \frac{P_B(q)P_B[\alpha(k,q,\mu)]}{\alpha^2(k,q,\mu)} \left[ 2k^5q^3\mu + k^4 q^4(1-5\mu^2) + 2k^3 q^5\mu^3\right]~,
\end{equation}
where $\alpha(k,q,\mu) = \sqrt{k^2+q^2-2kq\mu}$ and $\mu$ is the cosine of the angle between $\boldsymbol k$ and $\boldsymbol q$, that is $\boldsymbol k \cdot \boldsymbol q = kq\mu$. We note that, since the magnetic field power spectrum is affected by the ultraviolet cutoff at $k_\mathrm{max}$, the two integrals in the equation above actually do not cover the entire reported integration range. We are now in a position to estimate the total linear matter power spectrum and its evolution with cosmic time.

In Figure \ref{fig:spectrum} we show the contribution of magnetically induced density fluctuations to the total linear matter power spectrum, for different values of the magnetic spectral index ranging from $n_B = -2$ to $n_B = 2$. Here and in what follows, unless specified otherwise, we normalized the magnetic field power spectrum such that $\sigma_{B,0}(1~h^{-1}\mathrm{Mpc}) = 0.1$ nG. The generic effect of PMFs is to sharply increase the linear matter power spectrum at scales $k\sim 1-10~h$ Mpc$^{-1}$. On scales $k\gtrsim k_\mathrm{max}$ the impact of PMFs abruptly vanishes, due to the absence of magnetic field correlations above that wavenumber. In practice the magnetically induced linear power spectrum should be cut off at the magnetic Jeans scale $k_B$, which is significantly lower than $k_\mathrm{max}$ (see Eqs. \ref{eqn:kmax} and \ref{eqn:kb}). In Figure \ref{fig:spectrum} we did not include this cut-off, for demonstration reasons. However for the computation of the mass variance (Section \ref{sct:nonlinear} below) and all non-linear quantities that follow we did introduce the Jeans length cut-off. We finally note the fact that at large scales the slope of the power spectrum induced by PMFs is shallower for $n_B = -2$ than for larger spectral indices. This is expected, as in \cite{GO03.1} the authors shown that for $k/k_\mathrm{max}\ll 1$ the matter power spectrum is $\Pi(k)\propto k^4$ if $n_B > -3/2$, while $\Pi(k)\propto k^{2n_B+7}$ if $n_B < -3/2$. In the specific case of $n_B = -2$, we get $\Pi(k)\propto k^3$.

\section{Non-linear structures}\label{sct:nonlinear}

\subsection{Variance}\label{sct:variance}

\begin{figure}
\centering
\includegraphics[width=0.65\hsize]{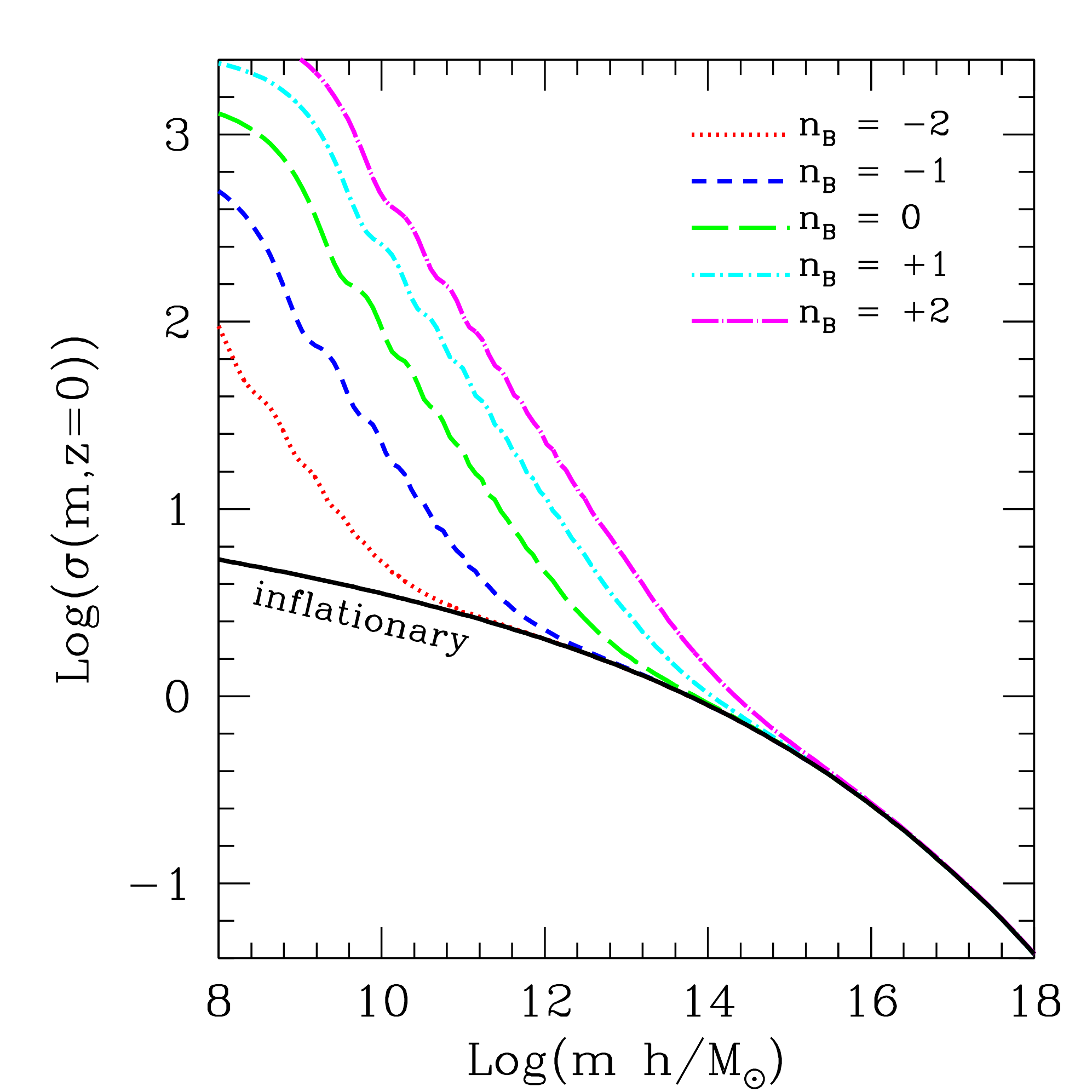}
\caption{The mean amplitude of the total matter density field smoothed on a scale corresponding to mass $m$, at $z = 0$. Different line styles refer to different slopes for the magnetic field power spectrum, while the amplitude is set by $\sigma_{B,0}(1~h^{-1}\mathrm{Mpc})$ = 0.1 nG. The black solid line shows the result obtained by ignoring PMFs.}
\label{fig:variance}
\end{figure}

An essential ingredient in order to estimate the mass function and bias of dark matter halos is the mass variance. In cosmological models with PMFs the variance has to be computed from the total linear matter power spectrum given in Eq. (\ref{eqn:spectrum}), that is

\begin{equation}
\sigma^2(R,z) = \frac{1}{2\pi^2}\int_0^{+\infty} k^2\mathrm{d}k P_\mathrm{m}(k,z) W^2(k,R)~,
\end{equation}
where $W(k,R)$ is the Fourier transform of a smoothing function, that we assumed to be a top-hat of radius $R$ in real space. In Figure \ref{fig:variance} we show the \emph{rms} (the square root of the variance) of the total density fluctuation field as a function of the mass corresponding to a given smoothing scale, at $z = 0$. Due to the small-scale bump in the matter power spectrum produced by PMFs, the variance approaches the inflationary value for high masses, while it tends to grow much faster than that and reach much larger values when moving to smaller and smaller masses. 

For the calculation of the variance, and for all applications that follow, we truncated the magnetically induced linear power spectrum at the magnetic Jeans wavenumber $k_B$. This is an approximation, since a complete treatment taking into account the backreaction of velocity fields for $k\sim k_B$ \cite{KI96.1} results in a growth factor that depends on scale and indeed vanishes for $k\gtrsim k_B$, but not with a sharp transition. Yet, this is a good approximation and has been used regularly in the literature \cite{SE05.2,TA11.1}. Because of this cut-off, in models with nonvanishing PMFs the variance (and hence the \emph{rms}) flattens out for masses smaller than that corresponding to the magnetic Jeans length. This magnetic Jeans mass can be estimated as

\begin{equation}
\left[\frac{M_B}{M(\lambda)}\right]^{5+n_B} \simeq 1.214\times 10^{-2}~h^6 \left[\frac{\sigma_{B,0}(\lambda)}{1~\mathrm{nG}}\right]^6 \left( \frac{\lambda}{1~h^{-1}\mathrm{Mpc}} \right)^{-6} \left( \frac{\Omega_{\mathrm{b},0}h^2}{0.023} \right)^{-3} \left( \frac{\Omega_{\mathrm{m},0}h^2}{0.135} \right)^{-3}~,
\end{equation}
where 

\begin{equation}
M(\lambda)\equiv \frac{\Omega_{\mathrm{m},0}H_0^2}{2G}\lambda^3 \simeq 3.161 \times 10^{11}h^{-1}M_\odot\left( \frac{\lambda}{1~h^{-1}\mathrm{Mpc}} \right)^3 \frac{\Omega_{\mathrm{m},0}}{0.272}
\end{equation}
is the average mass contained within the radius $\lambda$. For $\sigma_{B,0}(1~h^{-1}\mathrm{Mpc}) = 0.1$ nG and the slope values considered here the magnetic Jeans mass ranges between $M_B \sim 10^{9}-10^{10}~h^{-1}M_\odot$.

By looking at the curves that show the effect of PMFs in Figure \ref{fig:variance}, we note some oscillations in the mean density amplitude at low mass values. The reason for these are the oscillations present in the Fourier transform of the smoothing function, combined with the sharp cutoff of the magnetically induced matter power spectrum at the magnetic Jeans scale. As a matter of fact we verified that if the top-hat window function is replaced by a Gaussian (as many authors in the literature do, e.g. \cite{TA06.2}), the oscillations disappear. However, it should be kept in mind that virtually all the mass function theory is based on a top-hat window function, hence we chose to stick with this choice.

A few consideration are in order about the behavior of the mass variance described above. Since for high masses the variance is basically unaffected by the presence of PMFs, we expect the abundance of massive galaxy clusters not to be substantially modified, and if so, only in cases in which the amplitude or spectral index of the magnetic field power spectrum are large enough. On the other hand, one might argue that the abundance of low-mass dark matter halos ($m\lesssim 10^{12}~h^{-1}M_\odot$) should be largely increased. However, there is another effect that needs to be taken onto account. The mass function is exponentially sensitive to the mass variance only for the rarest objects at any given redshift. This means that, since at $z = 0$ low-mass objects are not in the exponential tail of the mass function, the impact of PMFs on their abundance is also expected to be relatively modest, and dictated only by the larger slope of the \emph{rms}.

\subsection{Halo Mass Function and Bias}\label{sct:bias}

\begin{figure}
\centering
\includegraphics[width=0.65\hsize]{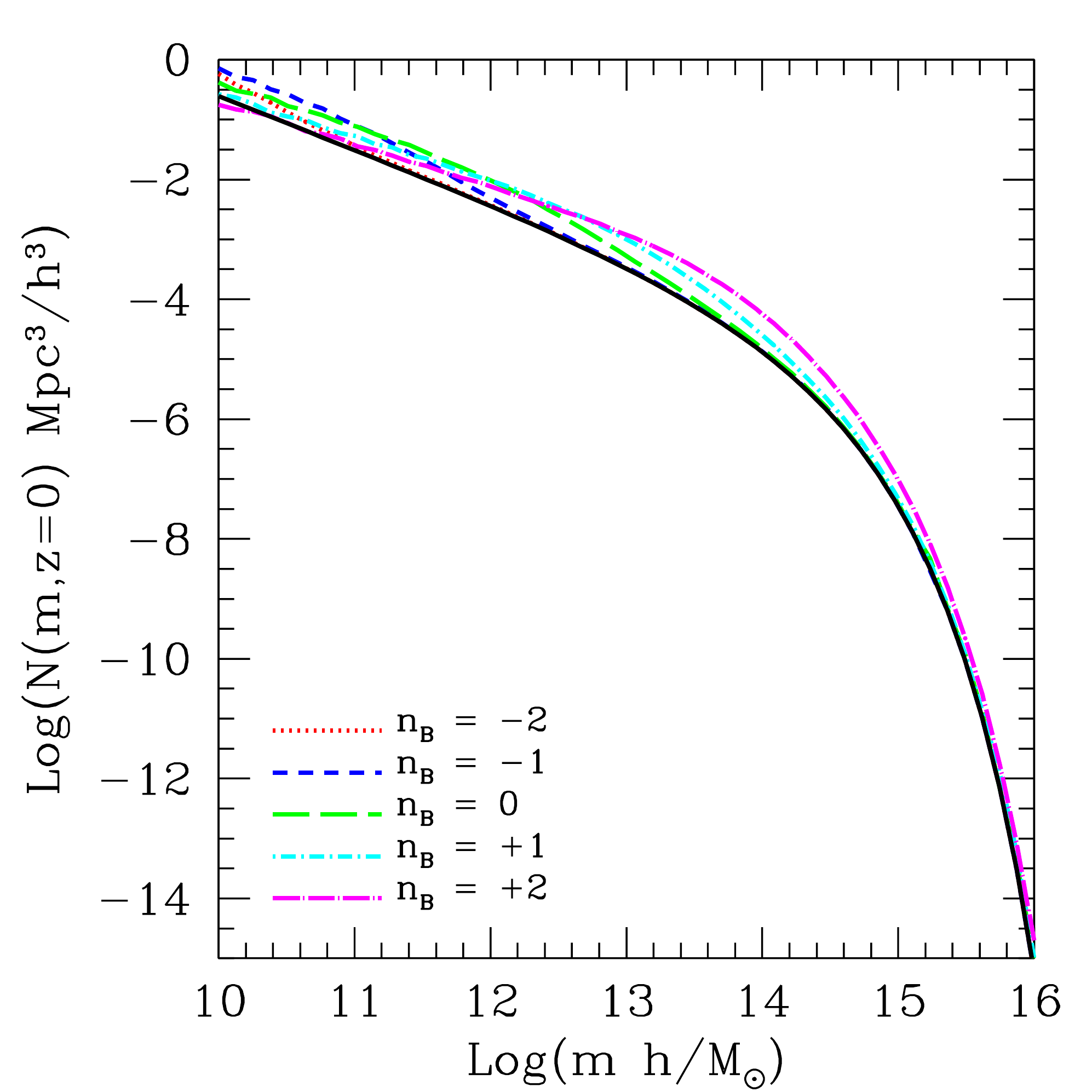}
\caption{The cumulative mass function for dark matter halos at $z = 0$, as a function of mass. The black solid line refers to the standard case with no PMFs, while the other lines refer to magnetic fields with different spectral slopes, as labeled. In all cases the magnetic field power spectrum is normalized by requiring that $\sigma_{B,0}(1~h^{-1}\mathrm{Mpc}) = 0.1$ nG.}
\label{fig:massFunction}
\end{figure}

We now turn attention to the estimates of dark matter halo mass function and bias. A proper treatment of these issues would need fully numerical cosmological simulations run with the modified initial conditions induced by PMFs. We plan to explore this topic in a future investigation, since this load of work is beyond the scope for the present paper. As a viable alternative, we adopted prescriptions based on semi-analytic considerations. Specifically, we employed the Sheth \& Tormen mass function \cite{SH02.1} and the Sheth, Mo, \& Tormen linear bias \cite{SH01.1} prescriptions. Since these formalisms are based at least partially on physical motivations (specifically, the ellipsoidal collapse model), we can argue them to remain acceptably valid in cosmologies including PMFs, as long as the mass variance $\sigma^2(m,z)$ is computed properly as described in Section \ref{sct:variance}.

\begin{figure}
\centering
\includegraphics[width=0.65\hsize]{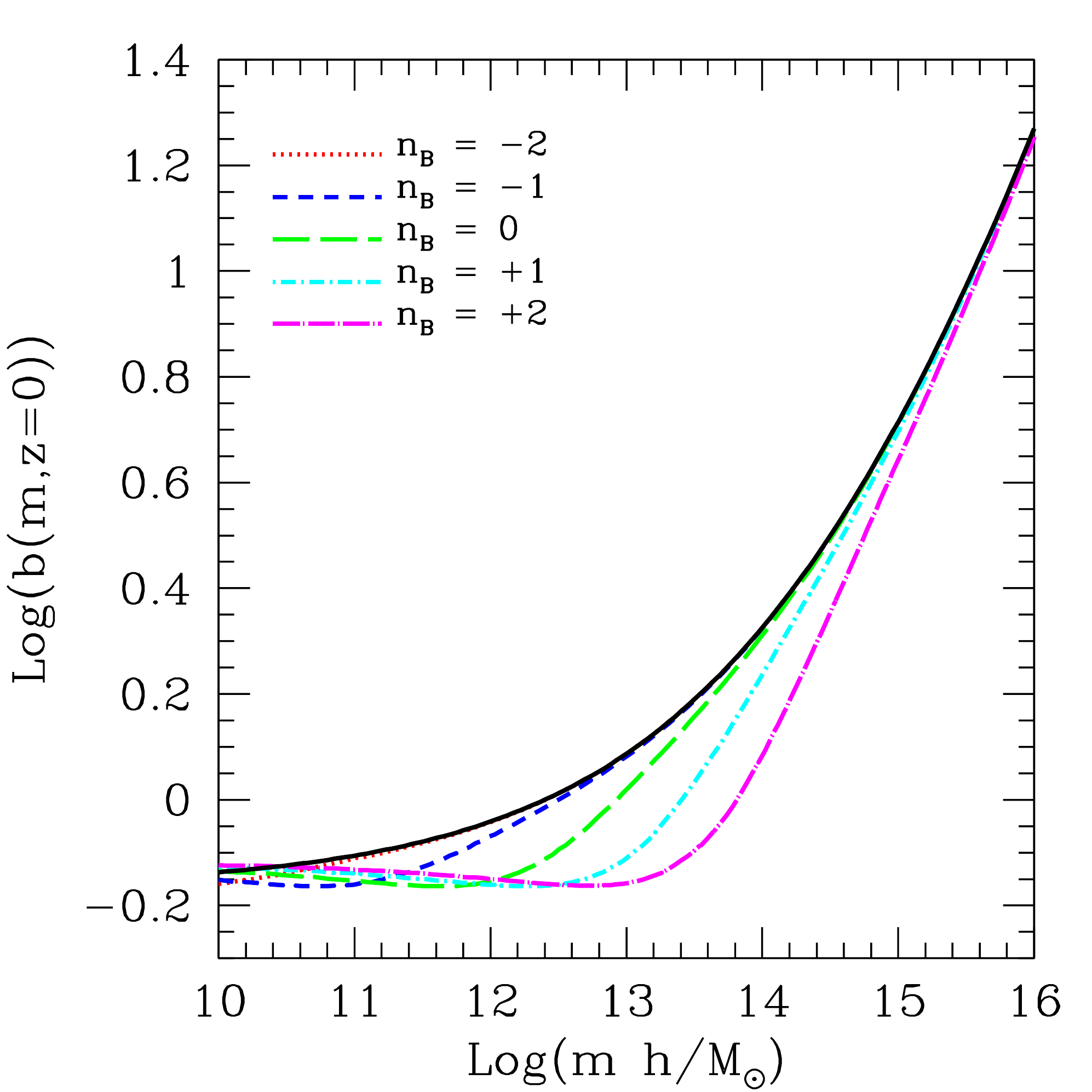}
\caption{The linear bias factor for dark matter halos at $z = 0$, as a function of mass. The black solid line refers to the standard case with no PMFs, while the other lines refer to magnetic fields with different spectral slopes, as labeled. In all cases the magnetic field power spectrum is normalized by requiring that $\sigma_{B,0}(1~h^{-1}\mathrm{Mpc}) = 0.1$ nG.}
\label{fig:bias}
\end{figure}

In Figure \ref{fig:massFunction} we show the mass function obtained with the Sheth \& Tormen prescription at $z=0$ for PMFs whose amplitude is set by $\sigma_{B,0}(1~h^{-1}\mathrm{Mpc}) = 0.1$ nG and for different spectral slopes. We plot the cumulative mass function, defined as

\begin{equation}
N(m,z) \equiv \int_m^{+\infty} \mathrm{d}\xi~n(\xi,z)~.
\end{equation}
As can be seen, the mass function is basically not modified at very high masses. This is due to the fact that at those masses the variance is also unaffected, because PMFs do not alter the linear matter power spectrum at large scales. The halo number counts get however increased by a factor of $\sim 5$ at intermediate masses, i.e., at $m \sim 10^{13} - 10^{14}~h^{-1} M_\odot$ for $n_B = 2$ and lower masses for lower magnetic spectral indices. At very low masses the effect of PMFs is again reduced (and eventually even reversed), due to the fact that those masses are not at the tip of the mass function at z = 0. Therefore their abundance is given by the slope of the variance as a function of mass (which flattens out below the magnetic Jeans mass) rather than the absolute value of the variance itself.

In Figure \ref{fig:bias} we report the linear bias of dark matter halos, computed according to the Sheth, Mo, \& Tormen prescription. In a fashion reminiscent of the mass function, the bias is almost unchanged at low and high masses, while it is substantially reduced (by up to $60\%$) at $m\sim 10^{13} - 10^{14}~h^{-1}M_\odot$ for $n_B = 2$. Also, the lower the magnetic spectral index, the lower the mass where the maximum reduction is observed, again similarly to the behavior of the mass function. This can be easily interpreted as a consequence of the increase in the variance shown in Figure \ref{fig:variance} being shifted at lower and lower masses as the magnetic spectral index decreases. Additionally, the fact that the halo bias drops in the same regime where the mass function increases is self-consistent, as more abundant objects have a lower clustering strength.

\subsection{Non-linear Power Spectrum}

In order to study cosmic shear, a model for the non-linear clustering of matter is needed. In this work we employed the halo model \cite{MA00.3,SE00.1,CO02.2}, a physically motivated framework that allows to write down the fully non-linear matter power spectrum as the sum of two terms, $P_\mathrm{NL}(k,z) = P_1(k,z)+P_2(k,z)$, where

\begin{equation}
P_1(k,z) = \int_0^{+\infty} \mathrm{d}m~n(m,z) \left[\frac{\hat\rho(m,z,k)}{\rho_{\mathrm{m},0}}\right]^2~,
\end{equation}
and

\begin{equation}
P_2(k,z) = P_m(k,z)\left[\int_0^{+\infty} \mathrm{d}m~n(m,z)~b(m,z) \frac{\hat\rho(m,z,k)}{\rho_{\mathrm{m},0}}\right]^2~.
\end{equation}
The first term defines the contribution from dark matter particle pairs belonging to the same halo, while the second term includes the clustering of different halos. Let us briefly review the ingredients of the model, while for details on the implementation we refer to \cite{FE09.4}. The function $b(m,z)$ represents the linear halo bias, while $n(m,z)$ is the halo mass function. We already described how both were modeled in Section \ref{sct:bias}. The function $\hat\rho(m,z,k)$ is the Fourier transform of the average dark matter density profile, $\rho(m,z,r)$, while $P_m(k,z)$ is the linear matter power spectrum, discussed in Section \ref{sct:linearspectrum}.

As the presence of PMFs introduces more matter power at small scales, it is likely that the internal structures of dark matter halos, and hence their average density profile, are modified with respect to the standard cosmological model without magnetic fields. Still, a full account of this modification could be obtained only with fully numerical cosmological simulations. Since these are not available yet, we assumed throughout that the average dark matter density profile is well represented by a NFW function \cite{NA96.1}. Moreover, we connected the concentration of the halo to its virial mass by adopting the model described in \cite{DO04.1}. This is admittedly an uncertainty of our approach, however in the subsequent cosmic shear analysis we limited ourselves to large and intermediate scales, so that the impact of the very inner structure of dark matter halos is not going to be relevant.

\begin{figure}
\centering
\includegraphics[width=0.65\hsize]{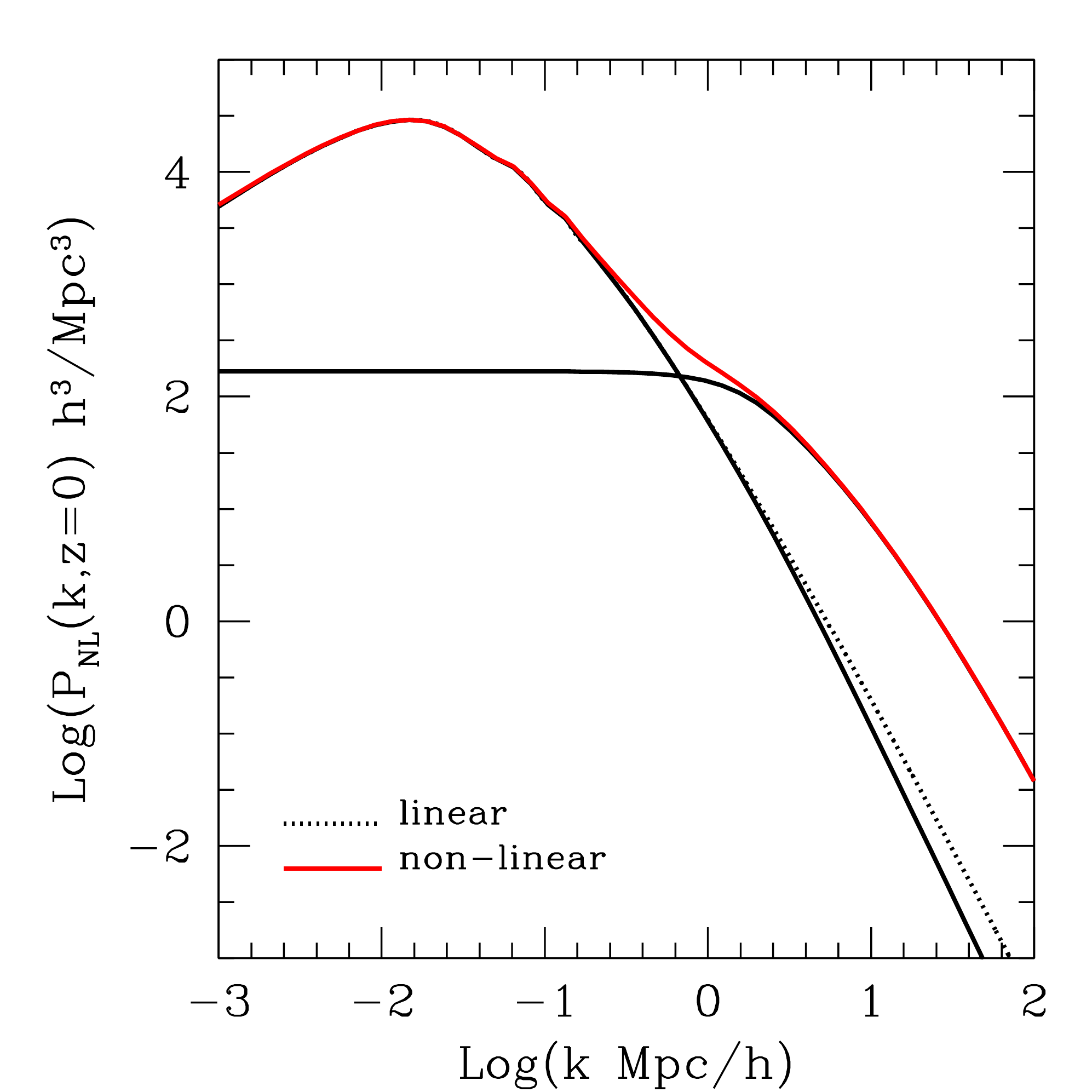}
\caption{The linear power spectrum and its non-linear counterpart obtained via the halo model at $z=0$, as labeled. PMFs are ignored here. The two black solid lines represent the $1-$halo (dominating at large $k$) and $2-$halo (dominating at small $k$) contributions to the non-linear spectrum.}
\label{fig:matterPower_regular}
\end{figure}

\begin{figure}
\centering
\includegraphics[width=0.65\hsize]{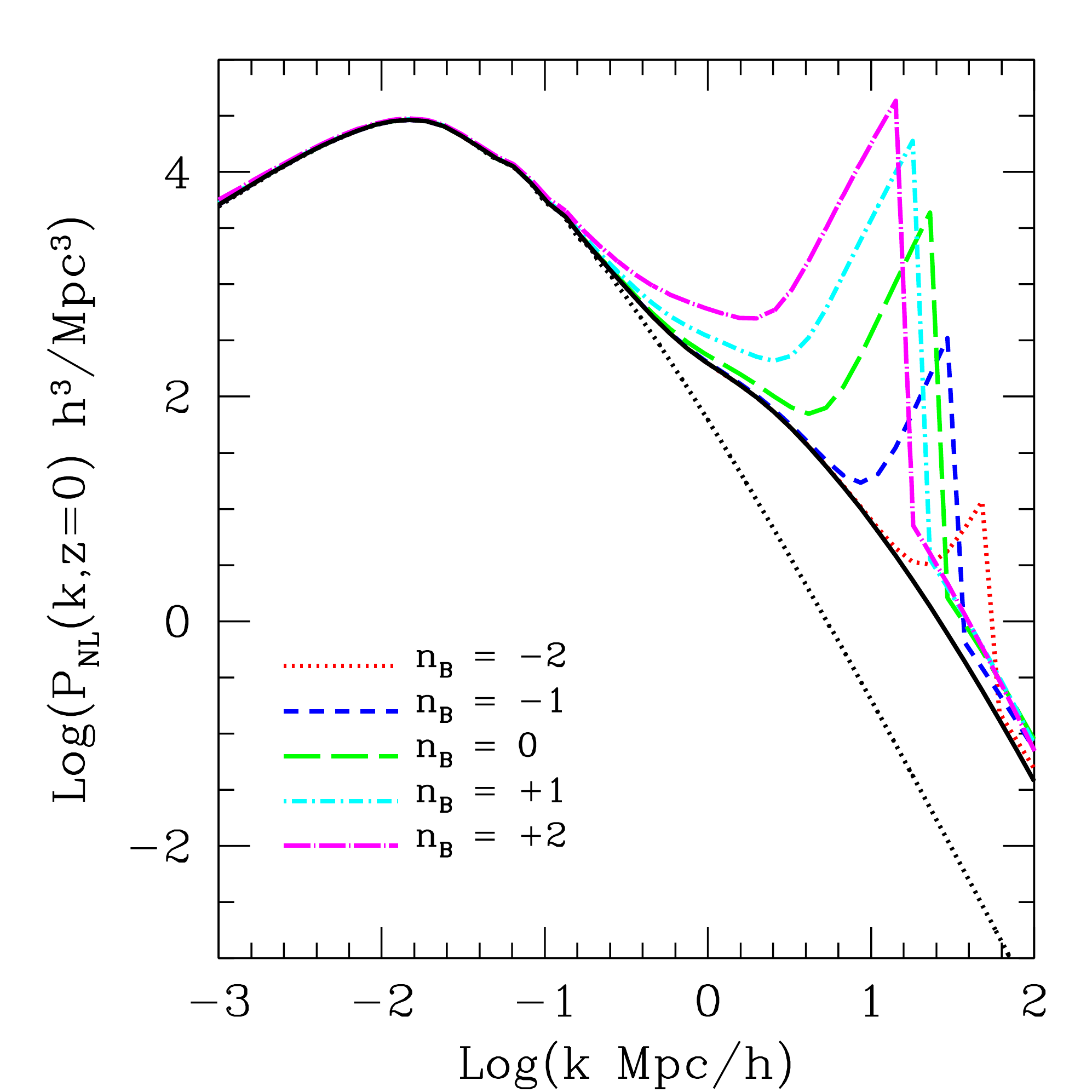}
\caption{The linear (black dotted line) and non-linear (black solid line) matter power spectra computed in absence of PMFs. The colored dashed lines represent the non-linear power spectra computed with PMFs having various spectral slopes, as labeled. In the latter cases the PMF amplitude is always set by requiring that $\sigma_{B,0}(1~h^{-1}\mathrm{Mpc}) = 0.1$~nG}
\label{fig:matterPower}
\end{figure}

In Figure \ref{fig:matterPower_regular} we compare the linear matter power spectrum in absence of PMFs with its non-linear counterpart obtained via the halo model. We also show the two contributions to the non-linear power spectrum, the $1-$halo term, dominating at small scales, and the $2-$halo term, dominating at large scales. One point worth noting is that, as well known, the $1-$halo term at large scales tends to a constant, namely

\begin{equation}\label{eqn:limit}
\lim_{k\to 0}P_1(k,z) = \int_0^{+\infty} \mathrm{d}m~n(m,z)~\frac{m^2}{\rho_{m,0}^2}~.
\end{equation}
Hence, the higher the mass function at high masses (note the $m^2$ weighting), the higher this constant. This is going to be relevant for the subsequent discussion.

In Figure \ref{fig:matterPower} we show how PMFs affect the non-linear matter power spectrum. In this plot the amplitude of PMFs is fixed, while results for different spectral indices are shown. As can be seen, the non-linear clustering of matter increases substantially at intermediate and small scales. One part of this increment is due to the corresponding increment that is seen in the linear matter power spectrum, shown in Figure \ref{fig:spectrum}. Please note that here we applied the sharp cutoff at the magnetic Jeans scale, so that the impact of PMFs disappears at smaller wavenumbers compared to Figure \ref{fig:spectrum}. Another part is more indirect, and shows up at scales larger than those affected by the former. This second part is due to the increase in the mass function at intermediate masses displayed in Figure \ref{fig:massFunction}, which results in an increase of the small-$k$ plateau of the $1-$halo term, as shown in Eq. (\ref{eqn:limit}). In other words, the $1-$halo term becomes important at larger scales as compared to the case with no PMFs, and this is more and more pronounced as the spectral index or the spectral amplitude become larger. 

The physical interpretation of this last fact is that structure formation starts earlier in cosmologies with PMFs than in the standard model, due to the enhanced mass variance that allows density perturbations to surpass the threshold for collapse at an earlier time. This means that density fluctuations with a given physical scale enter earlier in the non-linear regime or, turning the argument around, fluctuations that are still evolving linearly at a given redshift in the standard cosmology have already entered the non-linear evolution phase when PMFs are introduced. This finding is new and very important. Up to now the impact of PMFs on galaxy clustering has been ignored because it was argued that such an impact would manifest itself only at scales too small to be important. This is true for the linear power spectrum, but Figure \ref{fig:matterPower} shows that including non-linear modifications to the matter clustering can invalidate this assumption. More specifically, PMFs with large enough amplitude and/or spectral index may be able to affect the BAO region. This means that, in principle, a study of the BAO could be used to put constraints on cosmological magnetic fields. However this kind of study would require a more detailed modeling of the BAO scales, with either numerical simulations or perturbation theory. This is an interesting idea for future investigation.

\section{Cosmic shear}\label{sct:shear}

\subsection{Convergence power spectrum}

In order to derive the power spectrum of effective convergence \cite{BA01.1} from the fully non-linear matter power spectrum we adopted the Limber's approximation, which is valid for scales that are sufficiently small for the sky to be considered flat. Jeong \& Komatsu \cite{JE09.1} have shown that this approximation is good at the percent level for angular scales $\theta = 2\pi/\ell \lesssim 36$~degrees. Accordingly, the convergence power spectrum reads

\begin{equation}
C_\kappa(\ell) = \frac{9}{4}H_0^4\Omega_{\mathrm{m},0}^2\int_0^{\chi_\mathrm{h}} \mathrm{d}\chi \frac{W^2(\chi)}{a^2(\chi)} P\left[\frac{\ell}{f_K(\chi)},\chi \right]~.
\end{equation}
In the previous equation $\chi = \chi(z)$ is the comoving distance out to redshift $z$, the scale factor $a(\chi)$ is normalized such that $a(0)=1$, and $f_K(\chi)$ is the comoving angular diameter distance, which in general depends on the curvature of the Universe. In this work we considered only flat cosmological models, for which $f_K(\chi) = \chi$, however we keep the general notation for completeness.

The function $W(\chi)$ is a geometric weight, taking into account the fact that if the deflector is too close to the observer or to the sources the lensing strength is suppressed. It reads

\begin{equation}
W(\chi) = \int_0^{\chi_\mathrm{h}} \mathrm{d}\chi' g(\chi')\frac{f_K(\chi-\chi')}{f_K(\chi')}~,
\end{equation}
where $g(z)$ is the source redshift distribution. All these integrals formally extend up to the comoving horizon distance $\chi_\mathrm{h}$, however in practice the source redshift distribution drops to zero well before that distance is reached. In this work we adopted the source redshift distribution prescribed for \emph{Euclid} \cite{LA11.1}, adopting the configuration that covers $15,000$ square degrees. This is

\begin{equation}
g(z) = \frac{\beta}{z_\star~\Gamma\left[ (\alpha+1)/\beta\right]} \left( \frac{z}{z_\star} \right)^\alpha \exp\left[ -\left( \frac{z}{z_\star} \right)^\beta \right]~,
\end{equation}
where $\Gamma(a)$ is the complete Euler gamma function, $z_\star = 0.6374$, $\alpha = 2$, and $\beta=1.5$. The corresponding median source redshift is $z_\mathrm{m} = 0.9$.

\begin{figure*}
\centering
\includegraphics[width=0.65\hsize]{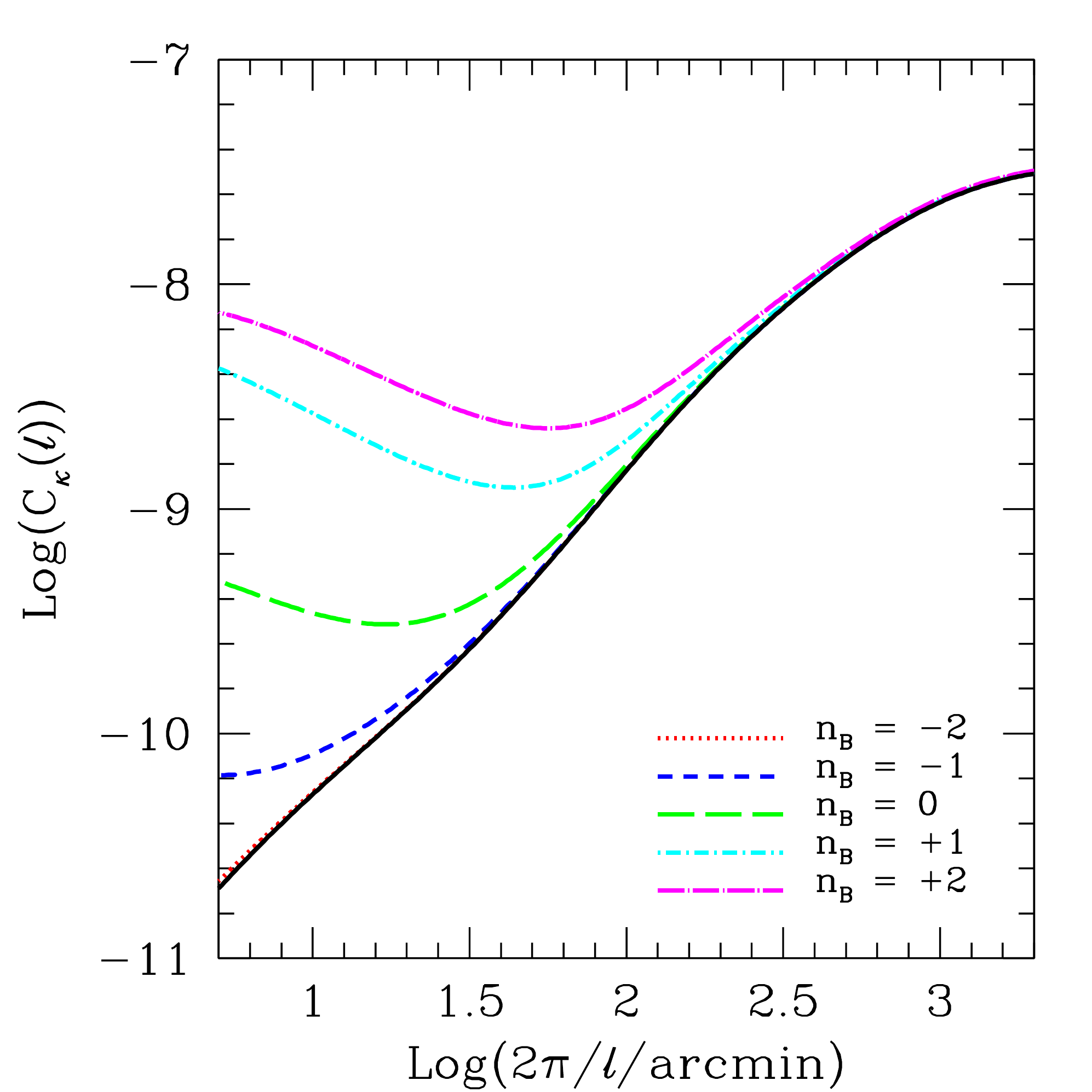}
\caption{The convergence power spectrum for the fiducial cosmology without magnetic fields (black solid line). The colored dashed lines refer to PMFs with different spectral slopes, as labeled. In all latter cases the magnetic field amplitude is set by requiring $\sigma_{B,0}(1~h^{-1}\mathrm{Mpc}) = 0.1$~nG.}
\label{fig:cosmicShear}
\end{figure*}

In Figure \ref{fig:cosmicShear} we show the convergence power spectrum computed for the various PMF spectral slopes considered in this work. As can be seen, the sharp cutoff corresponding to the magnetic Jeans scale is now smoothed out. This is due to the fact that a given angular scale in the sky corresponds to different physical scales at different redshifts, so that a given multipole gets contributions from physical scales both above and below the cutoff. We remind the reader that the comoving magnetic Jeans scale is independent of redshift \cite{KI96.1}. Consistently with the effect seen for the matter power spectrum, PMFs cause a substantial increase in convergence power at intermediate/small angular scales. For a fixed PMF amplitude, this effect is more marked for larger spectral slopes. This obviously implies that constraints on the spectral amplitude will be stronger for larger values of $n_B$.

\subsection{Constraints on PMFs}

In order to obtain bounds on the amplitude of PMFs it is necessary to quantify the statistical error on the convergence power spectrum estimated through cosmic shear. If ignoring tomography and the non-Gaussian part of weak lensing covariance, that however dominates only at small angular scales, such an error can be estimated as \cite{KA92.1,KA98.1,SE98.1,HU02.1}

\begin{equation}
\Delta C_\kappa(\ell) = \sqrt{\frac{2}{\Delta\ell(2\ell+\Delta\ell)f_\mathrm{sky}}} \left[ C_\kappa(\ell) + \frac{\sigma_\gamma^2}{\bar n} \right]~.
\end{equation}
In the previous equation $f_\mathrm{sky}$ represents the fraction of the sky covered by the weak lensing survey at hand, $\sigma_\gamma$ is the intrinsic shape contribution to the average galaxy ellipticity, and $\bar n$ is the average number density of background sources. The multipole interval $\Delta\ell$ is the width of multipole band where the power spectrum estimator is averaged upon. We adopted $\Delta\ell = 1$, and for the other parameters we chose the \emph{Euclid} specifications, that are $f_\mathrm{sky} = 0.364$, $\sigma_\gamma = 0.3$, and $\bar n = 30/$arcmin$^2$.

Given this, we can estimate a $\Delta\chi^2$ function by assuming that the measured (and hence true) convergence power spectrum is the one obtained in the standard model without PMFs, while the model power spectrum is the one obtained by including PMFs with a given amplitude and spectral index. Therefore we can write

\begin{equation}
\Delta\chi^2(A_{B,0},n_B) = \sum_{\ell=\ell_1}^{\ell_2} \left[\frac{C_\kappa(A_{B,0},n_B;\ell)-C^{(0)}_\kappa(\ell)}{\Delta C_\kappa^{(0)}(\ell)}\right]^2~,
\end{equation}
where $A_{B,0}$ is actually determined by $\sigma_{B,0}(1~h^{-1}\mathrm{Mpc})$ and the superscript $(0)$ refers to quantities evaluated within the fiducial framework having vanishing PMF amplitude. The choice of the multipole range where the sum is to be extended is important. The \emph{Euclid} specifications require $\ell_1 = 5$ and $\ell_2 = 5,000$, however we adopted two more conservative extrema. In particular, we set $\ell_1 = 50$, in order to make sure that the Limber's approximation is excellently satisfied, and we adopted $\ell_2 = 3,000$ in order to make sure that the impact of baryonic physics (e.g., \cite{SE11.1}), the non-Gaussian part of the weak lensing covariance, and the uncertainties on the inner structure of dark matter halos in the presence of PMFs can all be safely neglected. Needless to say, if all these uncertainties will be carefully modeled in the future, the analysis will be extended to substantially higher multipoles, with resulting tighter constraints on the amplitude of PMFs.

\begin{figure*}
\centering
\includegraphics[width=0.65\hsize]{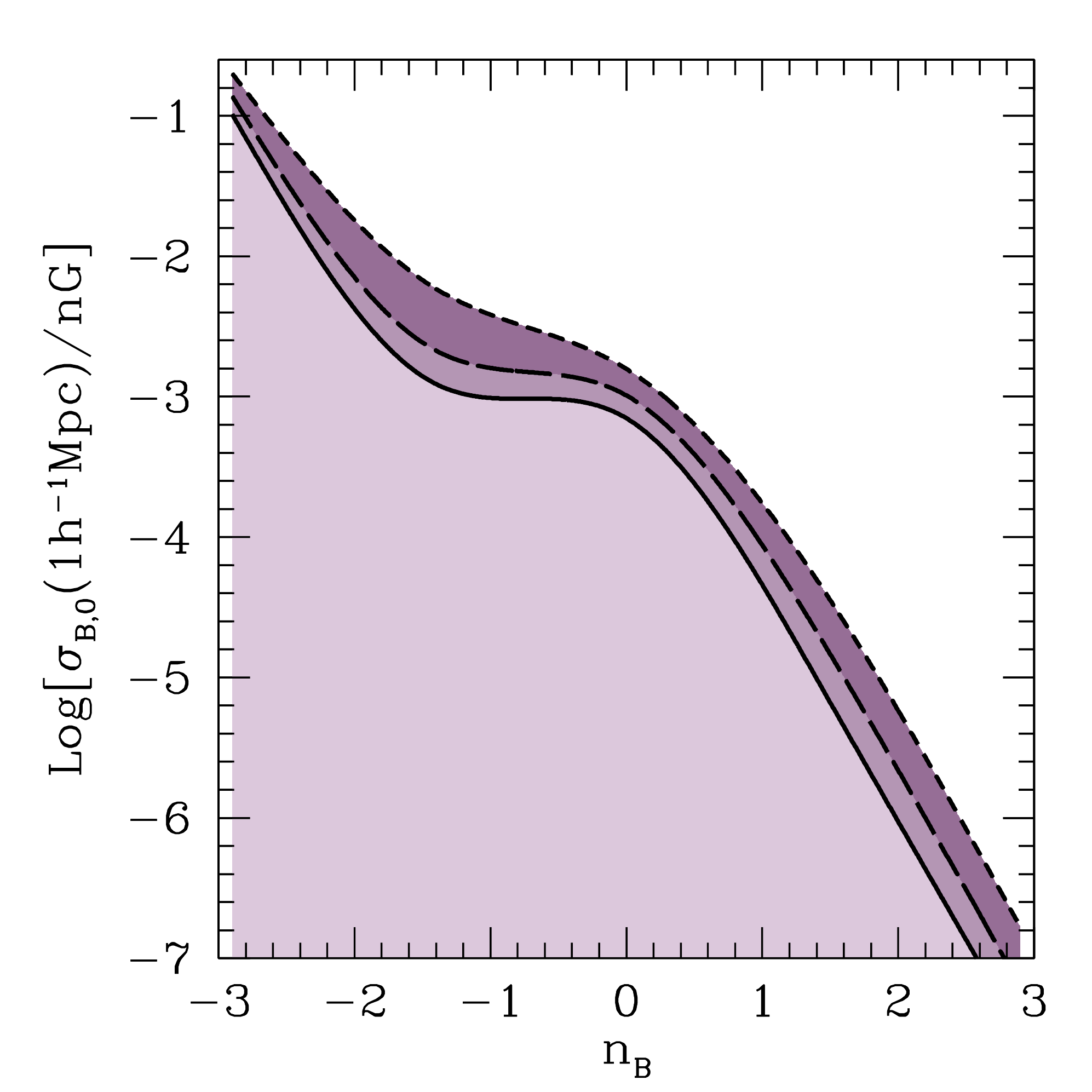}
\caption{Constraints (upper limits) on the PMF amplitude as a function of the spectral index $n_B$, obtained from cosmic shear. The different lines refer to different Confidence Levels (CLs). Specifically, the short dashed line shows the $99.7\%$ CL, the long dashed line refers to the $95.4\%$ CL, and the solid line displays the $68.3\%$ CL.}
\label{fig:bounds}
\end{figure*}

In Figure \ref{fig:bounds} we show the resulting constraints on the amplitude of PMFs as a function of the spectral index, for different confidence levels. We computed these bounds for a few selected values of $n_B$ and then interpolated over the full range of spectral indices shown. This procedure introduces a slight degree of uncertainty, however, given the very fluctuating nature of many of the present bounds on the amplitude of PMFs, and because the calculation of the convergence power spectrum in these models is quite time-consuming, we believe a full Monte-Carlo exploration of the parameter space is not warranted. As one might expect, the constraints are tighter for larger than for smaller spectral indices, because the impact of PMFs on the linear matter power spectrum is also larger in those cases (see Figure \ref{fig:spectrum}). Constraints on the amplitude of PMFs range from $\sim 0.1$ nG for spectral indices close to $n_B = -3$, down to $\sim 10^{-7}$ nG for $n_B \sim 3$. It is worth recalling that observations of secondary emission from gamma-ray sources place lower limits on the amplitude of PMFs at the level of $\sim 10^{-8} - 10^{-6}$ nG, thus implying that \emph{Euclid} will necessarily detect PMFs, if they exist and if they have a large spectral index, $n_B \sim 2-3$. The saddle point at $n_B\sim -1.5$ is interesting and likely due to the change in the slope of the magnetically induced matter power spectrum for that specific value of the spectral index.

\section{Summary and conclusions}\label{sct:conclusions}

In this work we investigated the impact of PMFs on cosmological weak lensing, and assessed the power of future wide-field cosmic shear surveys on the model of \emph{Euclid} for constraining the amplitude of these seed magnetic fields. \emph{Euclid} is a space mission selected for the ESA Cosmic Vision program, and it is currently scheduled for launch in $2019$. Our main results can be summarized as follows.

\begin{itemize}
\item Owing to the fact that PMFs induce a substantial increase in power at small scales, structure formation begins at earlier times if magnetic fields are taken into account. As a consequence, fluctuations on scales that at a given redshift are still in the linear regime for the standard cosmology with no magnetic fields, are already in the nonlinear evolutionary stage when PMFs are introduced.
\item Because of the anticipated non-linear evolution of density fluctuations and the increase in power at small scales, PMFs cause a substantial increment in the convergence power spectrum at intermediate and small angular scales. The increase is more marked for large spectral indices, because the power of matter fluctuations induced by PMFs is larger in those cases.
\item The former effect allows one to use cosmological weak lensing in order to put constraints on the amplitude of PMFs for a given spectral index. In agreement with the previous point, constraints are stronger for larger indices: adopting the \emph{Euclid} cosmic shear specifications results in bounds varying from $\sim 0.1$ nG for $n_B\sim -3$ down to $\sim 10^{-7}$ nG for $n_B\sim 3$. 
\end{itemize}

Our cosmic shear analysis relies on a series of simplifying assumptions. For instance, only the statistical errors on convergence power spectrum estimates were included, while weak lensing systematics such as photometric redshift errors and shape measurement uncertainties were neglected. Also, the degrading effect of uncertainties on the other cosmological parameters was not taken into account. For instance massive neutrinos, acting as a hot dark matter component, can wash away part of the small-scale power introduced by PMFs, and hence can somewhat worsen the constraints shown in Figure \ref{fig:bounds}. Furthermore, tomography, which has the potential of ameliorating the constraining power of weak lensing, was not implemented. These points can be improved upon in future investigations on PMFs, while for the sake of the present paper we were more interested in illustrating the approximate performance of cosmic shear statistics.

While this paper was being finalized we became aware of a similar study, performed in \cite{PA12.1}. There the authors used current cosmic shear data in order to constrain PMFs, by simply summing the PMF-induced matter power spectrum to the ordinary non-linear \cite{PE94.1} spectrum. The present work extends that by \cite{PA12.1} by including the impact of PMFs on the non-linear growth of cosmic structures. By comparing our results with theirs we can conclude that future cosmic shear surveys will improve constraints on PMF amplitude by about one order of magnitude compared to current data. To the best of our knowledge ours are the first forecasted constraints that are not based on the CMB, the primordial nucleosynthesis, or the linear matter power spectrum. For magnetic spectral indices $n_B \gtrsim 2$ the constraints from \emph{Euclid} will hit the lower limit set by the observations of secondary emission by gamma-ray sources, and hence in this case PMFs will be certainly detected, if they exist. It should however be recalled that bounds from the CMB and the gravitational wave background seem to favor a negative magnetic spectral index, although many uncertainties still remain in this respect. If this is the case, then PMFs will be detectable by \emph{Euclid} only if their amplitude is $\gtrsim 1$ pG.

It is worth recalling that the PMF amplitude and spectral index give important indications about the mechanism responsible for their formation, and of the epoch in the early Universe during which such formation occurred. For instance, PMF formation occurring during phase transitions is expected to produce a bluer spectrum as compared to inflationary formation \cite{YA12.1}. Thus, the study of PMFs by means of the LSS provides an exciting new window for studying the physics of the early Universe. Additionally, it will help to answer the fundamental question of the origin of astrophysical magnetic fields observed in galaxies and clusters of galaxies.

\section*{Acknowledgments}

CF acknowledges support from the University of Florida through the Theoretical Astrophysics Fellowship. LM is grateful for financial contributions from contracts ASI-INAF I/023/05/0, ASI-INAF I/088/06/0, ASI I/016/07/0 'COFIS', ASI 'Euclid-DUNE' I/064/08/0, ASI-Uni Bologna-Astronomy Dept. 'Euclid-NIS' I/039/10/0, and PRIN MIUR 'Dark energy and cosmology with large galaxy surveys'. We wish to thank an anonymous referee for comments and suggestions that helped improve the presentation of our work.

\bibliographystyle{JHEP}
\bibliography{master}

\end{document}